\documentclass[twocolumn,prl, preprintnumbers,amsmath,amssymb,superscriptaddress,showpacs]{revtex4}
\usepackage{natbib} 

\usepackage{graphicx}
\usepackage{dcolumn}
\usepackage{bm}

\begin{document}

\title{Entangled topological features of light}
\author{J. Romero}
\affiliation{Dept. of Physics and Astronomy, SUPA, University of Glasgow, Glasgow G12 8QQ UK}
\affiliation{Dept. of Physics, SUPA, University of Strathclyde, Glasgow G4 ONG UK}
\author{J. Leach}
\affiliation{Dept. of Physics and Astronomy, SUPA, University of Glasgow, Glasgow G12 8QQ UK}
\author{B. Jack}
\affiliation{Dept. of Physics and Astronomy, SUPA, University of Glasgow, Glasgow G12 8QQ UK}
\author{M.R. Dennis}
\affiliation{H.H. Wills Physics Laboratory,  University of Bristol, Bristol BS8 1TL UK}
\author{S. Franke-Arnold}
\affiliation{Dept. of Physics and Astronomy, SUPA, University of Glasgow, Glasgow G12 8QQ UK}
\author{S.M. Barnett} 
\affiliation{Dept. of Physics, SUPA, University of Strathclyde, Glasgow G4 ONG UK}
\author{M.J. Padgett}
\affiliation{Dept. of Physics and Astronomy, SUPA, University of Glasgow, Glasgow G12 8QQ UK}

\begin{abstract}We report the entanglement of topological features, namely, isolated, linked optical vortex loops in the light from spontaneous parametric down-conversion (SPDC). In three dimensions, optical vortices are lines of phase singularity and vortices of energy flow which percolate through all optical fields.  This example of entanglement is between features that extend over macroscopic and finite volumes, furthermore, topological features are robust to perturbation . The entanglement of photons in complex three-dimensional(3D) topological states suggests the possibility of entanglement of similar structures in other quantum systems describable by complex scalar functions, such as superconductors, superfluids  and Bose-Einstein condensates. 
\end{abstract}

\maketitle
Most  current  optical  experiments on entanglement  use SPDC  as  the  source  of correlated  photon  pairs. The simultaneous conservation of both energy and momentum in SPDC leads to various properties of the photons exhibiting quantum entanglement.  Correlations in their polarisation can be used to repeat the groundbreaking violations of the Bell inequality that were originally observed using a two-photon cascade source \cite{Freedman1972, A.Aspect1981}. Correlations in the spectra and arrival times of the down-converted photons demonstrate entanglement of energy and time\cite{AliKhan2006}. Correlations are also present in  their spatial modes \cite{Barreiro2005,Wagner2008} including those associated with orbital angular momentum (OAM) \cite{A.Mair2001}.  Optical eigenmodes of OAM have phase singularities (optical vortices) along their axis. These lines are defined as zeros of the complex field amplitude \cite{Jack2010}. As nodes, the phase on optical vortex lines is undefined (singular) and the phase changes by an integer multiple of $2\pi$ around a vortex line. 

Optical vortex lines can be found in optical speckle, familiar in light from a coherent laser beam scattered from a rough surface. Vortices occur between the bright speckles, and are seen as points where the 3D phase singularity lines intersect the viewing plane. In 3D speckle, these vortex lines form a fractal tangle on the large scale, percolating through space 
\cite{O'Holleran2008}, with many closed loops which are occassionally linked together \cite{O'Holleran2009}. Light produced by SPDC is spatially incoherent and can be considered as optical speckle. This is therefore a good candidate in which to observe topological vortex features\cite{Ghosh1986}. Here we show that quantum correlations apply to 3D vortex features of electromagnetic fields. Specifically, we show that the links of vortex loops, embedded within optical fields produced by SPDC are entangled.

\begin{figure}[h!]
\begin{center}
\includegraphics{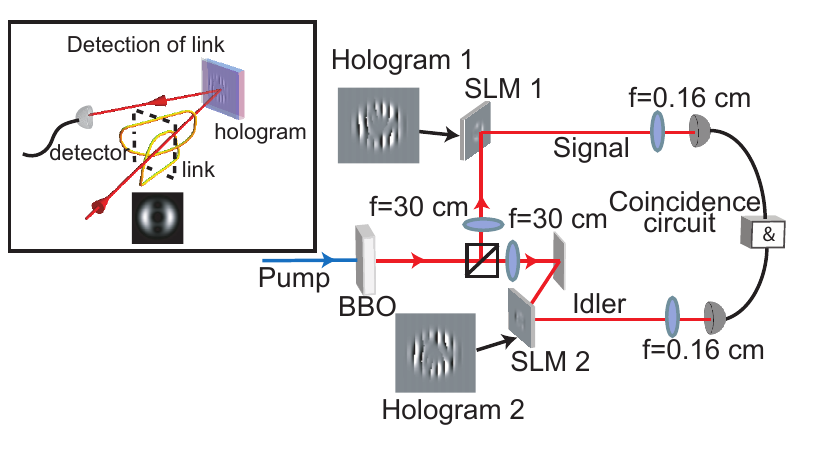}
\caption{\textbf{Schematic for measuring Hopf links.} The topological states of the signal and idler photons are measured by encoding by holograms displayed on separate spatial light modulators (SLMs).
}
\label{fig2}
\end{center}
\end{figure}

The Laguerre-Gaussian (LG) family of optical modes, as the Gaussian eigenmodes of OAM (and thus with axial vortex lines), is a convenient basis for specifying modal superpositions with linked vortex loops. 
LG modes have become important due to their easy experimental generation and simple mathematical representation; in particular, their OAM (and axial vortex) is characterized by $\exp(i\ell\phi)$, giving $\ell\hbar$ as the OAM per photon \cite{Allen1992}. The entanglement of the  OAM of photons has been observed experimentally \cite{A.Mair2001} and, more recently, has been demonstrated by violating the Bell inequality \cite{J.Leach2009}.  
However, the entanglement of OAM is not our primary concern here, but rather the entanglement of macroscopic vortex features that can be synthesised by combining LG modes, including those possessing no OAM.

The modal superpositions that form the links can be generated using specially designed diffractive optical components (holograms).  Although specified only in two dimensions, holograms determine the propagation of the 3D optical field behind them. Previously, holograms have transformed the Gaussian output of a laser or single mode fibre into a field  with a pair of linked phase singularity loops (Hopf links) \cite{J.Leach2005, Jack2010}. The fidelity of the intensity and phase produced by these holograms has been confirmed by tomographicaly imaging the phase and intensity structure of the holographically produced field \cite{Jack2010}. The same hologram that transforms a Gaussian mode into the Hopf link can be used in reverse as a measurement hologram, that is it can be used to detect the 3D field. In our case, this transforms a Hopf link back to the fundamental Gaussian mode which can then, and only then, be coupled to a single mode fibre and photon detector (fig. \ref{fig2} (inset)). A detection event constitutes the measurement of the 3D complex topological state. 

In modal superpositions of this kind, the linked vortex lines intertwine within regions of the field of very low optical intensity, and hence their topology is fragile to perturbation.  The practical generation and observation of these topological features relies on the numerical optimisation of the complex coefficients of the mode components to separate the vortex lines by regions of relatively high intensity \cite{Jack2010}.  Once the optimum coefficients of the modal components in the superposition are determined, it is a simple matter to design the corresponding detection hologram. 
\begin{figure}[h!]
\begin{center}
\includegraphics{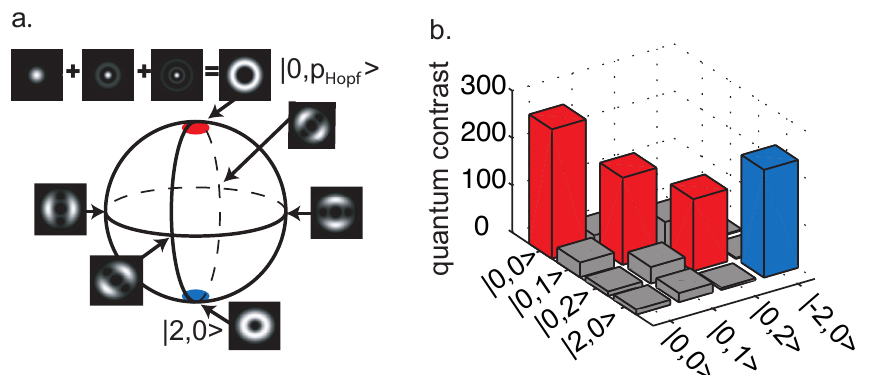}
\caption{\textbf{Bloch sphere for Hopf links and correlations among the constituent LG modes in the state (\ref{link}).} (a) A Bloch sphere for the Hopf links has the state $\left|0,p_{Hopf}\right\rangle$ (with constant phase and a small nonzero on-axis intensity) and  $\left|2,0\right\rangle$ (or $\left|-2,0\right\rangle$ ) at the poles. The equatorial states correspond to the Hopf links. (b)Red bars show the correlation between the $\ell=0$ states that make up $\left|0,p_{Hopf}\right\rangle$ (north pole) and the blue bar correspond to the correlation between the $\ell=2$ and $\ell=-2$ state (south pole). }
\label{fig3}
\end{center}
\end{figure}

The modal superposition we consider to produce the vortex Hopf link is given by,\begin{eqnarray}
\left|\Psi_{\rm Hopf\:link}\right\rangle&=& 0.264 \left|0,0\right\rangle - 0.628 \left|0,1\right\rangle \nonumber \\&+& 0.426 \left|0,2\right\rangle -0.596e^{i2\theta} \left|2,0\right\rangle.
\label{link}
\end{eqnarray}
where $\left|\ell,p\right\rangle$ denote the LG mode with $p$ radial nodes and azimuthal index $\ell$.
Ideally, for a plane wave pump beam of zero OAM $(\ell=0)$, the light from SPDC can be written in the LG basis as the entangled state,
\begin{equation}
\left|\Psi_{SPDC}\right\rangle=\sum^{\infty}_{0}\sum^{\infty}_{-\infty}c_{\ell,p}\left|\ell,p\right\rangle\left|-\ell,p\right\rangle
\label {Psi}
\end{equation}where $\left|c_{\ell,p}\right|^2$ is dependent on the down-conversion and is the probability of generating a photon pair in the $\left|\pm\ell,p\right\rangle$ state. The state (\ref{Psi}) has a wide range of different modes including the modes which comprise the superposition that could form the links. 
We can separate the superposition (\ref{link}) into two components consisting of modes with zero and nonzero OAM ($\ell=0$ and $\ell=2$ respectively), if we define
the state $\left|0,p_{Hopf}\right\rangle= 0.329 \left|0,0\right\rangle - 0.782 \left|0,1\right\rangle + 0.530 \left|0,2\right\rangle$, which we get by normalizing the first three terms of (\ref{link}). We can then write the Hopf link state as
$
\left|\Psi_{\rm Hopf\:link}\right\rangle=\alpha\left|0,p_{Hopf}\right\rangle -  \beta e^{i2\theta}\left|2,0\right\rangle,
$
where $\alpha=0.803$ and $\beta=0.596$. This is convenient because from the entire state (\ref{Psi}), we need have only to concern ourselves with the two-dimensional subspace given by
$
\left|\Psi_{L}\right\rangle= c_0\left|0,p_{Hopf}\right\rangle\left|0,p_{Hopf}\right\rangle+c_2\left|2,0\right\rangle\left|-2,0\right\rangle,
$
where $c_0$ and $c_2$ are complex coefficients dependent on the modal bandwidth of our down-conversion process, and in our case we measure to be 0.76 and 0.64 respectively. The advantage of a two-dimensional subspace is that it lends itself to traditional tests of entanglement such as the Bell inequality. Exploring this two-dimensional subspace allows us to investigate the possibility of entanglement between spatially separated 3D features.

The quantum mechanical prediction for the coincidence count rate $C$ of detecting the signal and idler photons in states $\left|\Psi_s\right\rangle$  and  $\left|\Psi_i\right\rangle$ respectively  then simplifies to the overlap integral between the complex amplitudes of the detected modes and the relevant modes of the pump beam $\left|\Psi_L\right\rangle$ \cite{Franke-Arnold2002},
\begin{equation}
C(\Psi_s,\Psi_i)\propto\left|\left\langle \Psi_s\right|\left\langle \Psi_i\right|\left.\Psi_L\right\rangle\right|^2
\label {C}
\end{equation}
In our experiment we normalise all the coincident counts to those which would be obtained from a system in which the signal and idler photons were uncorrelated.  We call this ratio the quantum contrast, given by $\rm QC=C/(S_iS_s\Delta t)$, where $\rm C$, $\rm S_i$ and $\rm S_s$ are the coincidence, idler and signal count rates respectively, and $\rm \Delta t$ is the timing resolution of our coincidence counting electronics.

It is well known that observing correlations in one measurement basis is not sufficient to show entanglement \cite{Mandel1995}.  At the heart of the mystery of entanglement are the correlations exhibited in the bases corresponding to incompatible observables, (e.g. momentum and position, linear and circular polarisation).  In a 2D state space, the concept of incompatible observables is best illustrated by a reference to a Bloch sphere where for example a rotation of linear polarisation is equivalent to a change in phase between the constituent circular polarisations. We can cast the measurement of Hopf links similarly, in that we can have a Bloch sphere with the states $\left|0,p_{Hopf}\right\rangle$ and $\left|0,2\right\rangle$ (or $\left|0,-2\right\rangle$ ) at the poles.  The equatorial states of this Bloch sphere are the Hopf links oriented at different angles (see fig. \ref{fig3}(a)). To show entanglement between the Hopf links then means demonstrating that the strength of nonlocal correlations depend not only on the magnitude of the modes (the poles of the sphere) but also on their relative phases. We test this phase dependence by changing both the relative axial $\Delta z$ positions and angular orientations $\Delta\theta$ of the topological features measured in the signal and idler beams.  Both of these shifts correspond to changes in phase of the modes comprising the superposition, arising from the Gouy phase and the rotational symmetry of the modes respectively \cite{Kawase2008}.  Rather than moving any of the optical components we apply these shifts directly simply by setting the phase of the modal superpositions from which the hologram design is derived.  The relative phase $\Delta\phi_k$ of each modal component $|\ell_k,p_k\rangle$ of the superposition is related to the axial displacement and angular orientations and is  given by $\Delta\phi_k=(\left|\ell_k\right|+2p_k+1)\tan^{-1}(\Delta z/z_R) + \ell_k\Delta\theta$, where $z_R$ is the Rayleigh range 
for the modes.  If the observed correlations were simply and solely due to classical conservation, then the strength of the correlations would show no phase dependence.

We employ the experimental configuration shown in fig. \ref{fig2}. A quasi-cw, mode-locked, UV pump beam at 355 nm  incident on a 3 mm long type-I BBO crystal.  The crystal is oriented in a collinear geometry such that the down-converted  710 nm signal and idler photons are both incident on the same beamsplitter.  The exit face of the crystal is imaged to separate spatial light modulators (SLM). The SLMs are used to display the holograms the measurement holograms which specify the links we aim to detect (inset).  These SLMs are reimaged to the input facets of single-mode fibres which are themselves coupled to single-photon detectors.  The coincidence count rate from the two detectors is recorded as the holograms displayed on the SLMs are updated. The gate time for this coincidence measurement is $10\:\rm ns$ and we obtain typical count rates of $200\: \rm s^{-1}$. The SLMs that we use are phase-only modulators, however, by appropriate design of an off-axis hologram they can be easily used to detect any modal superposition even those involving a modulation in intensity \cite{J.Leach2005}. The phase of the hologram,$\Phi_{\rm holo}$, is given by
$
\Phi_{\rm holo}(x,y)=\left[\Phi_{\rm link}(x,y) + \Phi_{\rm grating}(x,y)\right]_{\rm mod 2\pi} \times  {\rm sinc}^{2}[1-\pi I_{\rm link}(x,y)],
$
where $\Phi_{\rm link}(x,y)$ and $I_{\rm link}(x,y)$ refer to the phase and intensity cross sections of the link.  $\Phi_{\rm grating}(x,y)$ is the phase of the diffraction grating used to direct the light to the first order.        

Following (\ref{Psi}),in the image plane of the crystal, the signal and idler fields are complex conjugates of each other. This means if the photons are projected by the holograms which are encoded in states which are themselves complex conjugates of each other, the correlation between signal and idler beams should be high.  The LG basis that we use to describe our topological features is 
an orthonormal, complete set and consequently the correlation between any two modes of differing indices should, ideally, be zero.  Before examining the correlation between the topological features themselves we examine the measured correlations between the four LG modes that form the Hopf link (see fig.\ref{fig3}(b)).   As anticipated, the correlation between any mode and its complex conjugate is high, 
while its correlation with all other modes is low.  We note, however, that some nominally orthogonal modes have residual correlations.  This reflects the finite aperture of detection which means that the measured modes with the same $\ell$ (e.g.$\ell=0$ ) and different $p$  are not completely orthogonal.
\begin{figure}[h!]
\begin{center}
\includegraphics{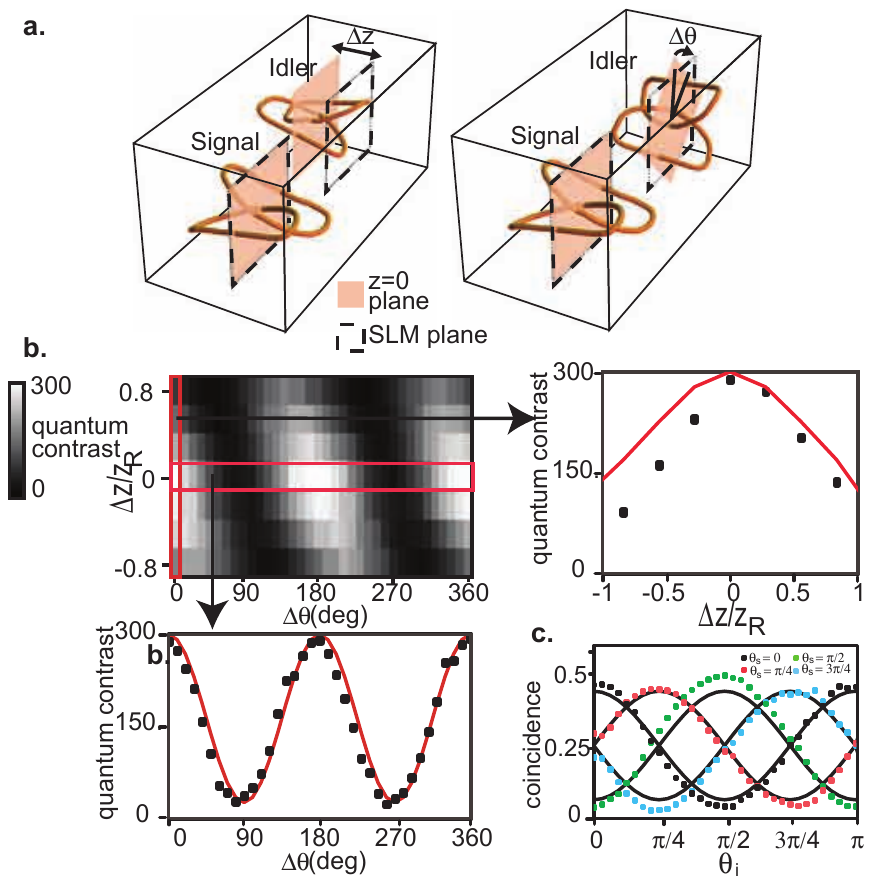}
\caption{\textbf{Experimental evidence of entanglement of topological states.} (a)Quantum contrast is measured for a Hopf link as we simultaneously vary both   axial and rotational displacements which are introduced by varying the hologram displayed on the SLM planes. (b) Coincidence measurements at various $\Delta z$ and $\Delta \theta$ reveal cross sections ( $\Delta z=0$ and $\Delta \theta=0$) that show the clear phase dependence. (c)Sinusoidal curves to show Bell inequality violation. }
\label{fig4}
\end{center}
\end{figure} 

We demonstrate the variation of the coincidence count rate between the Hopf links as measured in the signal and idler arms as the links are axially and rotationally displaced from each other. Fig. \ref{fig4}(b) is a plot of  the measured quantum contrast between a Hopf link and its phase conjugate as we simultaneously vary both the axial and rotational displacements in one of the arms. Fig. \ref{fig4}(b) also shows cross sections at $\Delta z=0$ and at $\Delta\theta=0$. The close agreement between experiment and that predicted by (\ref{C}) is a demonstration of the validity of quantum mechanics. 


The sinusoidal behavior of the coincidence count rate as we vary $\Delta\theta$ is reminiscent of the coincidence curves used to show a violation of the Bell inequality in the case of polarisation-entangled photons. Indeed, the fact that the state can be written in terms of two orthogonal set of modes suggests that we can perform a similar analysis for our Hopf links \cite{Bell1987}. 
We use the Clause-Horne-Shimony-Holt(CHSH) inequality \cite{Clauser1969}, an experimentally testable form of the Bell inequality derived to demonstrate the limit in the correlations possible with any local-realistic hidden variable theory. The inequality is given in terms of the Bell parameter S, and is violated when $\rm|S|>2$. The extent to which this inequality can be violated is an indication of the degree of entanglement of a quantum system, with S taking on a value of $2\sqrt{2}$ for maximally entangled states.

Following (\ref{C}), the predicted coincidence for orientations of the holograms $\theta_s$ and $\theta_i$ is given by
$
C(\theta_s,\theta_i)=c_0^2\alpha^4 + c_2^2\beta^4 + 2c_oc_2\alpha^2\beta^2\cos(2(\theta_s-\theta_i))=0.244+ 0.052 + 0.225\cos(2(\theta_s-\theta_i)).
$
Because the Hopf link contains a component with $\ell=0$, the Hopf link and its conjugate are not orthogonal hence the coincidence rate does not fall to zero as in the maximally entangled case. The maximum visibility expected in our case is 76\%, and S is 2.55.  Fig.\ref{fig4}(c) shows the coincidence curves used to calculate S from four different orientations of the signal hologram as the orientation of the idler hologram is varied from $0$ to $\pi$ (corresponding to a phase change between the $\left|0,p_{Hopf}\right\rangle$ and $\left|2,0\right\rangle$ of $0$ to 2$\theta_i=2\pi$).  The count rate for each $\theta_i$ setting is normalized such that the sum of the coincidence rate for each is unity.  The average visibility we obtain is 85\%, slightly higher than expected. This is a consequence of the interference arising from modes that are ideally orthgonal (i.e. the off-diagonal modes in fig.\ref{fig3}(b)), but experimentally gives residual coincidence counts. We measure S to be 2.44. This is lower than the maximum value of $2\sqrt{2}$, but is still greater than the classical limit of 2, thereby demonstrating that the Hopf links we measure in the down-converted fields are indeed entangled. 

The SLMs that we use have a diffraction efficiency of about 60\%. We are using off-axis holograms, meaning that any phase noise in the SLM will affect only the diffraction efficiency, not the phase of the measured state, which is instead set by the spatial form of the hologram.  The topology of the Hopf link has been shown to be robust against perturbations of the modal coefficients to as much as several percent. In any event, the fact that we violate the Bell inequality is an unambiguous demonstration of quantum entanglement and demonstrates that we are not constrained by the limitations of our SLMs.

Whereas earlier work in quantum entanglement has concentrated on point properties of the field (e.g. polarisation) or field cross sections (e.g. OAM), our work here concentrates on intrinsically three dimensional features of the field, we have demonstrated the existence of entangled topological features that occupy a finite \emph{volume}.
The wave description of light is equally applicable to other physical systems such as cold atoms and superfluids.  The existence of vortex lines and related topological features in these fields is an area of intense theoretical and experimental investigation \cite{Wal2000, Gullo2010,  Auslaender2009}. We postulate that the quantum entanglement of topological features of vortex lines may extend to cover these other system types. Moreover, the transfer of the topological vortex states from light to BEC \cite{Gullo2010,Kapale2005} may be a route to the preparation of macroscopically entangled topological states. In addition, topological states are usually robust and may therefore offer a route to increasing the stability of the entangled state\cite{Preskill1999}. 

We thank the UK EPSRC and Hamamatsu. MRD, SMB and MJP thank the Royal
Society and the Wolfson Foundation. We acknowledge the financial support of the Future and Emerging Technologies (FET) programme,HIDEAS number FP7-ICT-221906.

\bibliographystyle{plain}

\end{document}